\documentclass[12pt]{article}             
\usepackage{graphicx}

\begin{document}
\bibliographystyle{unsrt}
\setlength{\baselineskip}{18pt}
\parindent 24pt

\title{Quantum Entanglement and Quantum Discord \\of Two-Mode
Gaussian States\\ in a Thermal Environment
}
\author{Aurelian Isar \\Department of Theoretical Physics \\National Institute of Physics and Nuclear
Engineering\\
Bucharest-Magurele, P.O.Box MG-6, Romania\\
isar@theory.nipne.ro}

\date{}
\maketitle

\begin{abstract}
In the framework of the theory of open systems based on completely
positive quantum dynamical semigroups,
we give a description of the continuous-variable quantum entanglement and quantum discord for a system consisting of
two noninteracting modes embedded in a thermal environment.
Entanglement and discord are used to quantify the quantum correlations of the system. For all values of the temperature of the thermal reservoir, an initial separable Gaussian state remains separable for all times. We study the time evolution of logarithmic negativity, which characterizes the degree of entanglement, and show that in the case of an entangled initial Gaussian state, entanglement suppression (entanglement sudden death) takes place, for non-zero temperatures of the environment. Only for a zero temperature of the thermal bath the initial entangled state remains entangled for finite times.  We analyze the time evolution of the Gaussian quantum discord, which is a measure of all quantum
correlations in the bipartite state, including entanglement, and show that quantum
discord decays asymptotically in time under the effect of the thermal bath. This is contrast with the sudden death of
entanglement. Before the suppression of the entanglement, the qualitative evolution of quantum discord is very similar to that of the entanglement.
We describe also the time evolution of the degree of classical correlations and of quantum mutual information, which measures the total correlations of the quantum system.

\end{abstract}

\section{Introduction}

The study of quantum correlations is a key issue in quantum
information theory \cite{nie} and quantum entanglement represents the indispensable physical resource for the
description and performance of quantum information processing tasks,
like quantum teleportation, cryptography, superdense coding and
quantum computation \cite{hor1}. However, entanglement does not describe all the non-classical properties of
quantum correlations. For instance, recent theoretical and experimental results indicate that some non-entangled mixed states can improve
performance in some quantum computing tasks \cite{dat}. Zurek \cite{zur,oll} defined the quantum discord as a measure of quantum correlations
which includes entanglement of bipartite systems and it can also exist in separable states.
Very recently, quantum discord was related with the problem of irreversibility of entanglement \cite{fac} and an operational interpretation was given to quantum discord in terms on consumption of entanglement in an extended quantum state
merging protocol \cite{ani,cav}. The total amount of correlations contained in a quantum state is given by the quantum mutual information
which is equal to the sum of the quantum discord and
classical correlations \cite{hen}.

In recent years there is an increasing interest in using nonclassical entangled states of continuous variable systems in applications of quantum information processing, communication and computation \cite{bra1}. A full characterization of the nonclassical properties of such states exists, at present, only for the class of Gaussian states. In this special case there exist necessary and sufficient criteria of entanglement \cite{sim,dua} and quantitative entanglement measures \cite{vid,gie}. In quantum information theory of continuous variable systems, Gaussian states, in particular two-mode Gaussian states, play a key role since they can be easily created and controlled experimentally.

Implementation of quantum communication and computation encounters the difficulty that any realistic quantum system cannot be isolated and it always has to interact with its environment. Quantum coherence and entanglement of quantum systems are inevitably influenced during their interaction with the external environment. As a result of the irreversible and uncontrollable phenomenon of quantum decoherence, the purity and entanglement of
quantum states are in most cases degraded. Practically, compared with the discrete variable entangled states, the continuous variable entangled states may be more efficient because they are less affected by decoherence.

Due to the unavoidable interaction with the environment, any pure quantum state evolves into a mixed state and to describe realistically continuous variable quantum information processes it is necessary to take decoherence and dissipation into consideration. Decoherence and dynamics of quantum entanglement in continuous variable open systems have been intensively studied in the last years \cite{dua1,oli,ser3,pra,dod1,ser4,avd,ser2,ser1,ben1,ban,mch,man,jan,aphysa,aeur,arus1}.

In this work we study, in the framework of the theory of open systems based on completely positive quantum dynamical semigroups, the dynamics of the continuous variable quantum entanglement and quantum discord of a subsystem consisting of two uncoupled modes (harmonic oscillators) interacting with a common thermal environment. We
are interested in discussing the correlation effect of the
environment, therefore we assume that the two modes are
independent, i.e. they do not interact directly. The initial state of the subsystem is taken of Gaussian form and the evolution under the quantum dynamical semigroup assures the preservation in time of the Gaussian form of the state.
We have studied previously \cite{ascri,aosid} the evolution of the entanglement of two identical harmonic oscillators interacting with a general environment, characterized by general diffusion and dissipation coefficients. We obtained that, depending on the values of these coefficients, the state keeps for all times its initial type:
separable or entangled. In other cases, entanglement generation, entanglement sudden death or a periodic collapse and revival of entanglement take place.
In \cite{vas} it was analyzed the non-Markovian short-time-scale evolution of entanglement and quantum discord for initial two-mode squeezed thermal vacuum states of a system consisting of two identical and non-interacting harmonic oscillators coupled to either two independent bosonic baths or to a common bosonic bath. In the independent reservoirs case, it was observed the detrimental effect of the environment for these quantities. In the common reservoir case, for initial uncorrelated states, it was found that only quantum discord can be created via interaction with the bath, while entanglement remains absent.

The paper is organized as follows. In Sec. 2 we write the Markovian master equation in the Heisenberg representation for two uncoupled harmonic oscillators interacting with a general environment and give the general solution of the evolution equation for the covariance matrix,  i.e. we derive the variances and covariances of coordinates and momenta corresponding to a generic two-mode Gaussian state. By using the Peres-Simon necessary and sufficient condition for separability of two-mode Gaussian states \cite{sim,per}, we investigate in Sec. 3 the dynamics of entanglement for the considered subsystem. For all values of the temperature of the thermal reservoir, an initial separable Gaussian state remains separable for all times. We analyze the time evolution of the logarithmic negativity, which characterizes the degree of entanglement of the quantum state, and show that in the case of an entangled initial Gaussian state, entanglement suppression (entanglement sudden death) takes place, for non-zero temperatures of the environment. Only for a zero temperature of the thermal bath the initial entangled state remains entangled for all finite times, but in the limit of infinite time it evolves asymptotically to an equilibrium state which is always separable. We analyze the time evolution of the Gaussian quantum discord, which is a measure of all quantum
correlations in the bipartite state, including entanglement, and show that quantum
discord decays asymptotically in time under the effect of the thermal bath. This is contrast with the sudden death of
entanglement. Before the suppression of the entanglement, the qualitative evolution of quantum discord is very similar to that of the entanglement.
We describe also the time evolution of the degree of classical correlations and of quantum mutual information, which measures the total amount of correlations of the quantum system.
A summary is given in Sec. 4.

\section{Equations of motion of two independent modes interacting with the environment}

We study the dynamics
of the subsystem composed of two non-interacting modes in weak interaction with a thermal environment. In the axiomatic formalism
based on completely positive quantum dynamical semigroups, the irreversible time
evolution of an open system is described in the Heisenberg representation by the following quantum Markovian Kossakowski-Lindblad master equation
for an operator $A$ ($\dagger$ denotes Hermitian conjugation) \cite{lin,gor,rev}:
\begin{eqnarray}\frac{dA}{dt}=\frac{i}{\hbar}[H,A]+\frac{1}{\hbar}\sum_j(V_j^{\dagger}[A,
V_j]+[V_j^{\dagger},A]V_j).\label{masteq}\end{eqnarray}
Here, $H$ denotes the Hamiltonian of the open system
and the operators $V_j, V_j^\dagger,$ defined on the Hilbert space of $H,$
represent the interaction of the open system
with the environment.

We are interested in the set of Gaussian states, therefore we introduce such quantum
dynamical semigroups that preserve this set during time evolution of the system and in this case our model represents a Gaussian noise channel.
Consequently $H$ is
taken a polynomial of second degree in the coordinates
$x,y$ and momenta $p_x,p_y$ of the two quantum oscillators and
$V_j,V_j^{\dagger}$ are taken polynomials of first degree
in these canonical observables. Then in the linear space
spanned by the coordinates and momenta there exist only four
linearly independent operators $V_{j=1,2,3,4}$ \cite{san}: \begin{eqnarray}
V_j=a_{xj}p_x+a_{yj}p_y+b_{xj}x+b_{yj}y,\end{eqnarray} where
$a_{xj},a_{yj},b_{xj},b_{yj}$ are complex coefficients.
The Hamiltonian of the two uncoupled non-resonant harmonic
oscillators of identical mass $m$ and frequencies $\omega_1$ and $\omega_2$ is
\begin{eqnarray}
H={1\over 2m}(p_x^2+p_y^2)+\frac{m}{2}(\omega_1^2 x^2+\omega_2^2 y^2).\end{eqnarray}

The fact that the evolution is given by a dynamical semigroup
implies the positivity of the matrix formed by the
scalar products of the four vectors $ {\bf a}_x, {\bf b}_x,
{\bf a}_y, {\bf b}_y,$ whose entries are the components $a_{xj},b_{xj},a_{yj},b_{yj},$
respectively.
We take this matrix of the following form, where
all coefficients $D_{xx}, D_{xp_x},$... and $\lambda$ are real quantities (we
put from now on $\hbar=1$):
\begin{eqnarray} \left(\matrix{D_{xx}&- D_{xp_x} - i{\lambda}/{2}& D_{xy}&-
D_{xp_y} \cr - D_{xp_x} + i{\lambda}/{2}&D_{p_x p_x}&-
D_{yp_x}&D_{p_x p_y} \cr D_{xy}&- D_{y p_x}&D_{yy}&- D_{y p_y}
- i{\lambda}/{2} \cr - D_{xp_y} &D_{p_x p_y}&- D_{yp_y} + i{\lambda}/{2}&D_{p_y p_y}}\right).\label{coef} \end{eqnarray} It follows that
the principal minors of this matrix are positive or zero. From
the Cauchy-Schwarz inequality the following relations hold for the
coefficients defined in Eq. (\ref{coef}): \begin{eqnarray}
D_{xx}D_{p_xp_x}-D^2_{xp_x}\ge\frac{\lambda^2}{4},~
D_{yy}D_{p_yp_y}-D^2_{yp_y}\ge\frac{\lambda^2}{4},\nonumber\\
D_{xx}D_{yy}-D^2_{xy}\ge0,~
D_{p_xp_x}D_{p_yp_y}-D^2_{p_xp_y}\ge 0, \nonumber \\
D_{xx}D_{p_yp_y}-D^2_{xp_y}\ge 0,~D_{yy}D_{p_xp_x}-D^2_{yp_x}\ge 0.
\label{coefineq}\end{eqnarray}

We introduce the following $4\times 4$ bimodal covariance matrix:
\begin{eqnarray}\sigma(t)=\left(\matrix{\sigma_{xx}(t)&\sigma_{xp_x}(t) &\sigma_{xy}(t)&
\sigma_{xp_y}(t)\cr \sigma_{xp_x}(t)&\sigma_{p_xp_x}(t)&\sigma_{yp_x}(t)
&\sigma_{p_xp_y}(t)\cr \sigma_{xy}(t)&\sigma_{yp_x}(t)&\sigma_{yy}(t)
&\sigma_{yp_y}(t)\cr \sigma_{xp_y}(t)&\sigma_{p_xp_y}(t)&\sigma_{yp_y}(t)
&\sigma_{p_yp_y}(t)}\right).\label{covar} \end{eqnarray}
The problem of solving the master equation for the operators in Heisenberg representation can be transformed into a problem of solving first-order in time,
coupled linear differential equations for the covariance matrix elements.
Namely, from Eq. (\ref{masteq}) we obtain the following system of equations for the quantum correlations of the canonical observables ($\rm T$ denotes the transposed matrix) \cite{san}:
\begin{eqnarray}{d \sigma(t)\over
dt} = Y \sigma(t) + \sigma(t) Y^{\rm T}+2 D,\label{vareq}\end{eqnarray} where
\begin{eqnarray} Y=\left(\matrix{ -\lambda&1/m&0 &0\cr -m\omega_1^2&-\lambda&0&
0\cr 0&0&-\lambda&1/m \cr 0&0&-m\omega_2^2&-\lambda}\right),\\
D=\left(\matrix{
D_{xx}& D_{xp_x} &D_{xy}& D_{xp_y} \cr D_{xp_x}&D_{p_x p_x}&
D_{yp_x}&D_{p_x p_y} \cr D_{xy}& D_{y p_x}&D_{yy}& D_{y p_y}
\cr D_{xp_y} &D_{p_x p_y}& D_{yp_y} &D_{p_y p_y}} \right).\end{eqnarray}
The time-dependent
solution of Eq. (\ref{vareq}) is given by \cite{san}
\begin{eqnarray}\sigma(t)= M(t)[\sigma(0)-\sigma(\infty)] M^{\rm
T}(t)+\sigma(\infty),\label{covart}\end{eqnarray} where the matrix $M(t)=\exp(Yt)$ has to fulfill
the condition $\lim_{t\to\infty} M(t) = 0.$
In order that this limit exists, $Y$ must only have eigenvalues
with negative real parts. The values at infinity are obtained
from the equation \begin{eqnarray}
Y\sigma(\infty)+\sigma(\infty) Y^{\rm T}=-2 D.\label{covarinf}\end{eqnarray}

\section{Dynamics of continuous variable entanglement and discord}

To describe the dynamics of quantum correlations, we use two types of measures:
logarithmic negativity for entanglement, and quantum discord.

\subsection{Time evolution of entanglement and logarithmic negativity}

A well-known sufficient condition for inseparability is the
so-called Peres-Horodecki criterion \cite{per,hor}, which is based on
the observation that the non-completely positive nature of the
partial transposition operation of the density matrix for a bipartite system (transposition with respect to degrees of freedom of one subsystem only) may turn an inseparable state
into a nonphysical state. The signature of this non-physicality,
and thus of quantum entanglement, is the appearance of a negative
eigenvalue in the eigenspectrum of the partially transposed
density matrix of a bipartite system. The characterization of the separability
of continuous variable states using second-order moments of quadrature operators was given in Refs. \cite{sim,dua}. For Gaussian states,
whose statistical properties are fully characterized by just
second-order moments, this criterion was proven to be necessary
and sufficient: A Gaussian continuous variable state is separable if and only if the partial transpose
of its density matrix is non-negative (positive partial
transpose (PPT) criterion).

The two-mode Gaussian state is entirely specified by its
covariance matrix (\ref{covar}), which is a real,
symmetric and positive matrix with the following block
structure:
\begin{eqnarray}
\sigma(t)=\left(\begin{array}{cc}A&C\\
C^{\rm T}&B \end{array}\right),\label{cm}
\end{eqnarray}
where $A$, $B$ and $C$ are $2\times 2$ Hermitian matrices. $A$
and $B$ denote the symmetric covariance matrices for the
individual reduced one-mode states, while the matrix $C$
contains the cross-correlations between modes. When these correlations
have non-zero values, then the states with $\det C\ge 0$ are
separable states, but for $\det C <0$ it may be possible that
the states are entangled.

The $4\times 4$ covariance matrix (\ref{cm}) (where all first moments
can be set to zero by means of local unitary operations which do not affect the entanglement) contains
four local symplectic invariants in form of the determinants
of the block matrices $A, B, C$ and covariance matrix $\sigma.$ Based on the above invariants,
Simon \cite{sim} derived the following PPT criterion for bipartite Gaussian
continuous variable states: the necessary and sufficient condition for separability is
$S(t)\ge 0,$ where \begin{eqnarray} S(t)\equiv\det A \det B+(\frac{1}{4} -|\det
C|)^2\nonumber\\
- {\rm Tr}[AJCJBJC^{\rm T}J]- \frac{1}{4}(\det A+\det B)
\label{sim1}\end{eqnarray} and $J$ is the $2\times 2$ symplectic matrix
\begin{eqnarray}
J=\left(\begin{array}{cc}0&1\\
-1&0\end{array}\right).
\end{eqnarray}

We suppose that the asymptotic state of the considered open system is a Gibbs state corresponding to two independent quantum harmonic oscillators in thermal equilibrium at temperature $T.$ Then the quantum diffusion coefficients have the following form \cite{rev}:
\begin{eqnarray}m\omega_1 D_{xx}=\frac{D_{p_xp_x}}{m\omega_1}=\frac{\lambda}{2}\coth\frac{\omega_1}{2kT},\nonumber\\
m\omega_2 D_{yy}=\frac{D_{p_yp_y}}{m\omega_2}=\frac{\lambda}{2}\coth\frac{\omega_2}{2kT},\label{envcoe}\\
D_{xp_x}=D_{yp_y}=D_{xy}=D_{p_xp_y}=D_{xp_y}=D_{yp_x}=0.\nonumber\end{eqnarray}

The elements of the covariance matrix can be
calculated from Eqs. (\ref{covart}), (\ref{covarinf}). Solving for the time evolution of
the covariance matrix elements, we can obtain the entanglement
dynamics by using the Simon criterion.

For Gaussian states, the measures of entanglement of bipartite systems are based on some invariants constructed from the elements of the covariance matrix \cite{oli,ser4,avd}. In order to quantify the degree of entanglement of the infinite-dimensional bipartite system states of the two oscillators it is suitable to use the logarithmic negativity.  For a Gaussian density operator, the logarithmic negativity is completely defined by the symplectic spectrum of the partial transpose of the covariance matrix. It is given by
$
E_N=-\log_2 2\tilde\nu_-,
$
where $\tilde\nu_-$ is the smallest of the two symplectic eigenvalues of the partial transpose $\tilde{{\sigma}}$ of the 2-mode covariance matrix $\sigma$ \cite{ser4}:
\begin{eqnarray}2\tilde{\nu}_{\mp}^2 = \tilde{\Delta}\mp\sqrt{\tilde{\Delta}^2
-4\det\sigma}
\end{eqnarray}
and $ \tilde\Delta$ is the symplectic invariant (seralian), given by
$ \tilde\Delta=\det A+\det B-2\det C.$

In our model, the logarithmic negativity is calculated as \begin{eqnarray}E_N(t)=-\frac{1}{2}\log_2[4g(\sigma(t))], \end{eqnarray} where \begin{eqnarray}g(\sigma(t))=\frac{1}{2}(\det A +\det
B)-\det C\nonumber\\
-\left({\left[\frac{1}{2}(\det A+\det B)-\det
C\right]^2-\det\sigma(t)}\right)^{1/2}.\end{eqnarray}
It determines the strength of entanglement for $E_N(t)>0,$ and if $E_N(t)\le 0,$ then the state is
separable.

In the following, we analyze the dependence of the Simon function $S(t)$  and of the logarithmic negativity $E_N(t)$ on time $t$ and temperature $T$ of the thermal bath, with the diffusion coefficients given by Eqs. (\ref{envcoe}). We consider two types of the initial Gaussian states: 1) separable and 2) entangled.

1) We consider a separable initial Gaussian state, with the two modes initially prepared in their single-mode squeezed states (unimodal squeezed state) and with its initial covariance matrix taken of the form
\begin{eqnarray}\sigma_s(0)=\frac{1}{2}\left(\matrix{\cosh r&\sinh r &0&0\cr
\sinh r&\cosh r&0&0\cr
0&0&\cosh r&\sinh r\cr
0&0&\sinh r&\cosh r}\right),\label{ini1} \end{eqnarray}
where $r$ denotes the squeezing parameter. In this case $S(t)$ becomes strictly positive after the initial moment of time ($S(0)=0),$ so that the initial separable state remains separable for all values of the temperature $T$ and for all times.

2) The evolution of an entangled initial state is illustrated in Figure 1, where we represent the dependence of the logarithmic negativity $E_N(t)$ on time $t$ and temperature $T$ for an entangled initial Gaussian state, taken of the form of a two-mode vacuum squeezed state, with the initial covariance matrix given by
\begin{eqnarray}\sigma_e(0)=\frac{1}{2}\left(\matrix{\cosh r&0&\sinh r &0\cr
0&\cosh r&0&-\sinh r\cr
\sinh r&0&\cosh r&0\cr
0&-\sinh r&0&\cosh r}\right).\label{ini2} \end{eqnarray}
We observe that for a non-zero temperature $T,$ at certain finite moment of time, which depends on $T,$ $E_N(t)$ becomes zero and therefore the state becomes separable. This is the so-called phenomenon of entanglement sudden death. It is in contrast to the quantum decoherence, during which the loss of quantum coherence is usually gradual \cite{aphysa,arus}. For $T=0,$ $E_N(t)$ remains strictly positive for finite times and tends asymptotically to 0 for $t\to \infty.$ Therefore, only for zero temperature of the thermal bath the initial entangled state remains entangled for finite times and this state tends asymptotically to a separable one for infinitely large time. One can also show that the dissipation favorizes the phenomenon of entanglement sudden death -- with increasing the dissipation parameter $\lambda,$ the entanglement suppression happens earlier.

The dynamics of entanglement of the two oscillators depends strongly on the initial states and the coefficients describing the interaction of the system with the thermal environment (dissipation constant and temperature). As expected, the logarithmic negativity has a behaviour similar to that one of the Simon function in what concerns the characteristics of the state of being separable or entangled \cite{ascri,aosid,arus,aijqi}.

\begin{figure}
{
\includegraphics{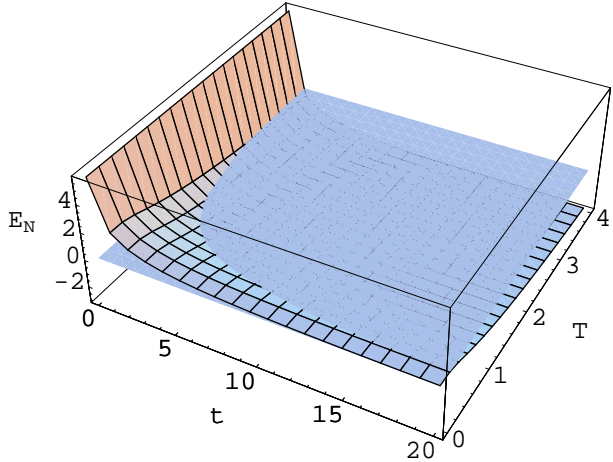}
}
\caption{Logarithmic negativity $E_N$ versus time $t$ and temperature $T$ for an entangled initial vacuum squeezed state with squeezing parameter $r=4$ and $\lambda=0.1, \omega_1=\omega_2=1.$ We take $m=\hbar=k=1.$
}
\label{fig:1}
\end{figure}

\subsection{Asymptotic entanglement}

On general grounds, one expects that the effects of
decoherence is dominant
in the long-time regime, so that no quantum correlations (entanglement) is expected to be left at infinity.
Indeed, using the diffusion coefficients given by Eqs. (\ref{envcoe}), we obtain from Eq. (\ref{covarinf}) the
following elements of the asymptotic matrices $A(\infty)$ and $B(\infty):$
\begin{eqnarray} m\omega_1\sigma_{xx}(\infty)=\frac{\sigma_{p_xp_x}(\infty)}{m\omega_1}=\frac{1}{2}\coth\frac{\omega_1}{2kT}, ~~~\sigma_{xp_x}(\infty)=0,\nonumber\\
m\omega_2\sigma_{yy}(\infty)=\frac{\sigma_{p_yp_y}(\infty)}{m\omega_2}=\frac{1}{2}\coth\frac{\omega_2}{2kT}, ~~~\sigma_{yp_y}(\infty)=0
\label{varinf} \end{eqnarray}
and of the entanglement matrix $C(\infty):$
\begin{eqnarray}\sigma_{xy} (\infty) =
\sigma_{xp_y}(\infty)=
\sigma_{yp_x}(\infty)=\sigma_{p_xp_y} (\infty) =0.\end{eqnarray}
Then the Simon expression (\ref{sim1}) takes the following form in the limit of large times: \begin{eqnarray} S(\infty)=
\frac{1}{16}\left(\coth^2\frac{\omega_1}{2kT}-1\right)\left(\coth^2\frac{\omega_2}{2kT}-1\right),\label{sim2}\end{eqnarray} and, correspondingly, the equilibrium asymptotic state is always separable in the case of two non-interacting harmonic oscillators immersed in a common thermal reservoir.

In Refs. \cite{ascri,aosid,arus,aijqi,arus2,ascri1} we described the dependence of the logarithmic negativity $E_N(t)$ on time and mixed diffusion coefficient for two harmonic oscillators interacting with a general environment. In the present case of a thermal bath, the asymptotic logarithmic negativity is given by (for $\omega_1\le\omega_2$)
\begin{eqnarray} E_N(\infty)=-\log_2\coth\frac{\omega_2}{2kT}.\end{eqnarray}
It depends only on temperature, and does not depend on the initial Gaussian state. $E_N(\infty)<0$ for $T\neq 0$ and $E_N(\infty)=0$ for $T=0,$ and this confirms the previous statement that the asymptotic state is always separable.

\subsection{Gaussian quantum discord}

The separability of quantum states has often been described as a property synonymous with the classicality.  However, recent studies have shown that separable states, usually
considered  as being classically correlated, might  also contain quantum correlations.
Quantum discord was introduced \cite{zur,oll} as a measure of all quantum
correlations in a bipartite state, including -- but not restricted to -- entanglement.
Quantum discord has been defined as the difference
between two quantum analogues of classically equivalent expression of
the mutual information, which is a measure of total  correlations in a quantum state. For pure entangled states quantum
discord coincides with the entropy of entanglement.  Quantum
discord can be different from zero also for some mixed separable
state and therefore the
correlations in such separable states with positive discord are an indicator of quantumness.
States with zero discord represent essentially a classical probability distribution embedded in a quantum system.

For an arbitrary bipartite state
$\rho_{12},$ the total correlations are expressed by quantum mutual information \cite{gro}
\begin{eqnarray}
I(\rho_{12})=\sum_{i=1,2} S(\rho_{i})-S(\rho_{12}),
\end{eqnarray} where $\rho_i$ represents the reduced density matrix of subsystem $i$ and $S(\rho)= - {\rm Tr}(\rho \ln \rho)$
is the von Neumann entropy. Henderson and Vedral \cite{hen} proposed a measure of bipartite classical
correlations $C(\rho_{12})$ based on a complete set of local projectors $\{\Pi_{2}^k\}$ on the subsystem 2: the classical correlation in the bipartite quantum state $\rho_{12}$ can be given by
\begin{eqnarray}
C(\rho_{12})=S(\rho_{1})-{\inf}_{\{\Pi_{2}^k\}}\{S(\rho_{1|2})\},
\end{eqnarray}
where $S(\rho_{1|2}) =\sum_{k}p^k S(\rho_{1}^k)$
is the conditional entropy of subsystem {1} and $\inf\{S(\rho_{1|2})\}$
represents the minimal value of the entropy with respect to a complete set of local measurements $\{\Pi_{2}^k\}.$
Here, $p^k$ is the measurement probability for the $k$th local projector and $\rho_{1}^k$ denotes the reduced state of subsystem $1$ after the local measurements.
Then the quantum discord is defined by
\begin{eqnarray}
D(\rho_{12})=I(\rho_{12})-C(\rho_{12}).
\end{eqnarray}

Originally the quantum discord was defined and evaluated mainly for finite dimensional systems. Very recently \cite{par,ade} the notion of discord has been extended to the domain of
continuous variable systems, in particular to the analysis of
bipartite systems described by two-mode Gaussian states.
Closed formulas have been derived for bipartite thermal squeezed states \cite{par} and for all two-mode Gaussian states \cite{ade}.

The Gaussian quantum discord of a general two-mode Gaussian state $\rho_{12}$
can be defined as the quantum discord where the conditional entropy is restricted to generalized Gaussian
positive operator valued measurements (POVM) on the mode 2 and in terms of symplectic invariants it is given by (the symmetry between the two modes 1 and 2 is broken) \cite{ade}
\begin{eqnarray}
D=f(\sqrt{\beta})-f(\nu_-) - f(\nu_+) + f(\sqrt{\varepsilon}),
\label{disc}
\end{eqnarray}
where \begin{eqnarray}f(x) =\frac{x+1}{2} \log\frac{x+1}{2} -\frac{x-1}{2} \log\frac{x-1}{2},\end{eqnarray}
\begin{eqnarray}\label{infdet}
\varepsilon=
 &  &
\hspace*{-.5cm}
\left\{  \hspace*{-.5cm}  \begin{array}{rcl}& &\begin{array}{c}\displaystyle{\frac{{2 \gamma^2+(\beta-1)(\delta-\alpha)
+2 |\gamma| \sqrt{\gamma^2+(\beta-1) (\delta-\alpha)}}}{{(\beta-1){}^2}}}\end{array},\\& &\qquad
\hbox{if}~~(\delta-\alpha\beta)^2 \le (\beta+1)\gamma^2 (\alpha +\delta)\\ \\& &
\begin{array}{c}\displaystyle{\frac{{\alpha\beta-\gamma^2+\delta-\sqrt{\gamma^4+(\delta-\alpha\beta){}^2-
2\gamma^2(\delta+\alpha\beta)}}}{{2\beta}}}\end{array}, \\& & \qquad
\hbox{otherwise,} \end{array} \right.
\end{eqnarray}
\begin{eqnarray}\alpha=4\det A,~~~\beta=4\det B,~~~\gamma=4\det C,~~~\delta=16\det\sigma,\end{eqnarray}
and $\nu_\mp$ are the symplectic eigenvalues of the state, given by
\begin{eqnarray}2{\nu}_{\mp}^2 ={\Delta}\mp\sqrt{{\Delta}^2
-4\det\sigma},
\end{eqnarray}
where
$\Delta=\det A+\det B+2\det C.$
Notice that Gaussian quantum discord only depends on $|\det C|$, i.e., entangled ($\det C<0$) and separable states are treated on equal footing.

\begin{figure}
{
\includegraphics{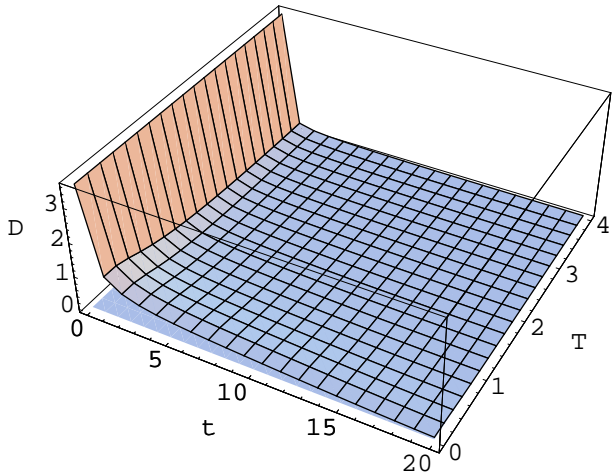}
}
\caption{Gaussian quantum discord $D$ versus time $t$ and temperature $T$ for an entangled initial vacuum squeezed state with squeezing parameter $r=4$ and $\lambda=0.1, \omega_1=\omega_2=1.$ We take $m=\hbar=k=1.$
}
\label{fig:2}
\end{figure}

The evolution of the Gaussian quantum discord $D$ is illustrated in Figure 2, where we represent the dependence of $D$ on time $t$ and temperature $T$ for an entangled initial Gaussian state, taken of the form of a two-mode vacuum squeezed state (\ref{ini2}), for such values of the parameters which satisfy for all times the first condition in formula (\ref{infdet}). The Gaussian discord has nonzero values for all finite times and this fact certifies the existence of nonclassical correlations in two-mode Gaussian states -- either separable or entangled. Gaussian discord asymptotically decreases in time, compared to the case of the logarithmic negativity, which has an evolution leading to a sudden suppression of the entanglement. For entangled initial states the Gaussian discord remains strictly positive in time and in the limit of infinite time it tends asymptotically to zero, corresponding to the thermal product (separable) state, with no correlation at all. One can easily show that for a separable initial Gaussian state with covariance matrix (\ref{ini1}) the quantum discord is zero and it keeps this values during the whole time evolution of the state.

From Figures 1 and 2 we notice that, in concordance with the general properties of the Gaussian quantum discord \cite{ade}, the states can be either separable or entangled for $D\le 1$ and all the states above the threshold $D=1$ are entangled.
We also notice that the decay of quantum discord is stronger when the temperature $T$ is increasing.

\begin{figure}
{
\includegraphics{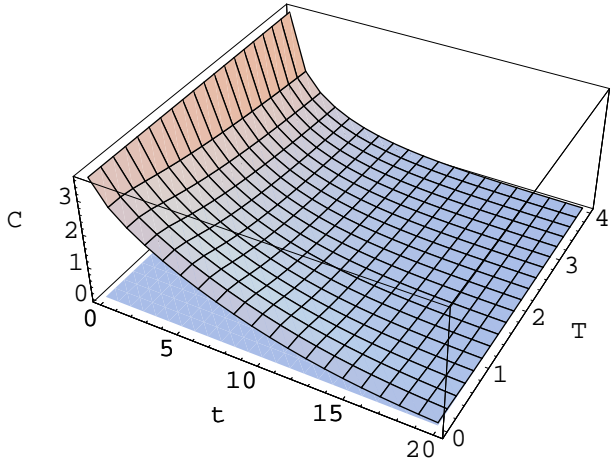}
}
\caption{Degree of classical correlations $C$ versus time $t$ and temperature $T$ for an entangled initial vacuum squeezed state with squeezing parameter $r=4$ and $\lambda=0.1, \omega_1=\omega_2=1.$ We take $m=\hbar=k=1.$
}
\label{fig:3}
\end{figure}

It should be remarked
that the decay of quantum discord is very similar to that of the entanglement before the time of
the sudden death of entanglement. In the vicinity of a zero logarithmic negativity ($E_N = 0$), the nonzero values of the discord can quantify the nonclassical correlations for separable
mixed states and one considers that this fact could make possible some tasks in quantum computation \cite{yut}.

The measure of classical correlations for a general two-mode Gaussian state $\rho_{12}$
can also be calculated and it is given by \cite{ade}
\begin{eqnarray}
C=f(\sqrt{\alpha}) - f(\sqrt{\varepsilon}),
\label{clas}
\end{eqnarray}
while the expression of the quantum mutual information, which measures the total correlations, is given by
\begin{eqnarray}
I=f(\sqrt{\alpha}) + f(\sqrt{\beta}) -f(\nu_-) - f(\nu_+).
\label{mut}
\end{eqnarray}

In Figures 3 and 4 we illustrate the evolution of the degree of classical correlations $C$ and, respectively, of the quantum mutual information $I,$ as functions of time $t$ and temperature $T$ for an entangled initial Gaussian state, taken of the form of a two-mode vacuum squeezed state (\ref{ini2}). These two quantities manifest a qualitative behaviour similar to that one of the Gaussian discord: they have nonzero values for all finite times and in the limit of infinite time they tend asymptotically to zero, corresponding to the thermal product (separable) state, with no correlation at all.
One can also see that the classical correlations and quantum mutual information decrease with increasing the temperature of the thermal bath.

\begin{figure}
{
\includegraphics{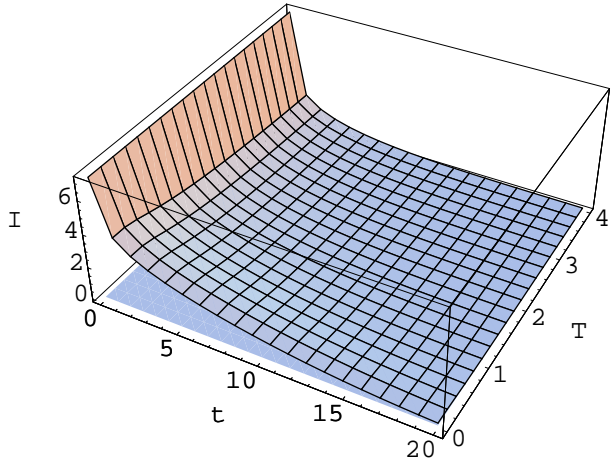}
}
\caption{Quantum mutual information $I$ versus time $t$ and temperature $T$ for an entangled initial vacuum squeezed state with squeezing parameter $r=4$ and $\lambda=0.1, \omega_1=\omega_2=1.$ We take $m=\hbar=k=1.$
}
\label{fig:4}
\end{figure}

\section{Summary}

In the framework of the theory of open systems based on completely positive quantum dynamical semigroups, we investigated the Markovian dynamics of quantum correlations for a subsystem
composed of two noninteracting modes embedded in a thermal bath. We have presented and discussed the influence of the environment on the dynamics of quantum entanglement and quantum discord for different initial states.
We have described the time evolution of the logarithmic negativity, which characterizes the degree of entanglement of the quantum state, in terms
of the covariance matrix for Gaussian input
states, for the case when the asymptotic state of the considered open system is a Gibbs state corresponding to two independent quantum harmonic oscillators in thermal equilibrium. The dynamics of the quantum entanglement strongly depends on the initial states and the parameters characterizing the environment (dissipation coefficient and temperature). For all values of the temperature of the thermal reservoir, an initial separable Gaussian state remains separable for all times. In the case of an entangled initial Gaussian state, entanglement suppression (entanglement sudden death) takes place for non-zero temperatures of the environment. Only for a zero temperature of the thermal bath the initial entangled state remains entangled for finite times, but in the limit of infinite time it evolves asymptotically to an equilibrium state which is always separable. The time when the entanglement is suppressed, decreases with increasing the temperature and dissipation.

We described also the time evolution of the Gaussian quantum discord, which is a measure of all quantum
correlations in the bipartite state, including entanglement.
The values of quantum discord decrease asymptotically in time. This phenomenon is different from the sudden death of entanglement. The time evolution
of quantum discord is very similar to that of the entanglement
before the sudden suppression of the entanglement. After the sudden death of the entanglement, the nonzero values of quantum discord
manifest the existence of quantum correlation for separable mixed states.  Quantum discord
is decreasing with increasing the temperature. One considers that the robustness of quantum discord could favorize the realization of scalable quantum computing in contrast to the fragility of the entanglement \cite{yut}.

Presently there is a large debate relative to the physical interpretation existing behind the fascinating phenomena of quantum decoherence and existence of quantum correlations - quantum entanglement and quantum discord. Due to the increased interest manifested towards
the continuous variables approach \cite{bra1,bra} to quantum information theory,
the present results, in particular the existence of quantum discord and the possibility of maintaining a bipartite entanglement in a
thermal environment for long
times, might be useful in controlling entanglement and discord in open systems and also for applications in the field of quantum information
processing and communication.

\section*{Acknowledgments}

The author acknowledges the financial support received from the Romanian Ministry of Education and Research, through
the Projects IDEI 497/2009 and PN 09 37 01 02/2010.

\end{document}